\documentclass[letterpaper,twocolumn,10pt]{article}
\usepackage[table]{xcolor}
\definecolor{red}{HTML}{ea5545}
\definecolor{mypink}{HTML}{f46a9b}
\definecolor{orange}{HTML}{ef9b20}
\definecolor{mymustardyellow}{HTML}{edbf33}
\definecolor{yello}{HTML}{ede15b}
\definecolor{myolivegreen}{HTML}{bdcf32}
\definecolor{green}{HTML}{87bc45}
\definecolor{blue}{HTML}{27aeef}
\definecolor{purple}{HTML}{b33dc6}

\usepackage{usenix-2020-09}

\usepackage{tikz}
\usepackage{amsmath}

\usepackage{filecontents}

\usepackage{listings} 
\usepackage{graphicx}
\usepackage{tcolorbox}
\usepackage{multicol}
\usepackage{xspace}
\usepackage{tabularx}
\usepackage{booktabs}
\usepackage{graphicx}
\usepackage{multirow}
\usepackage{hyperref}
\usepackage{xurl}
\usepackage{tabularx} 
\usepackage{fancyhdr} 

\newcommand{\mytab}{Tab.\@~}
\newcommand{\mytablong}{Table\@~}
\newcommand{\mysec}{Sec.\@~}
\newcommand{\myseclong}{Section\@~}
\newcommand{\myfig}{Fig.\@~}
\newcommand{\myfiglong}{Figure~}

\newcommand{\papername}{\textsc{Memory DisOrder}\xspace}
\newcommand{\papernameshort}{\textsc{DisOrder}\xspace}

\newcommand*\circledBlack[1]{\tikz[baseline=-0.3em]{
            \node[shape=circle,draw,inner sep=2pt, scale=0.6, fill=black, text=white, font=\normalsize, font=\bfseries] (char) {#1};}}
            
\newcommand*\circledWhite[1]{\tikz[baseline=-0.3em]{
            \node[shape=circle,draw,inner sep=2pt, scale=0.6, fill=white, text=black, font=\normalsize, font=\bfseries] (char) {#1};}}

\newcolumntype{L}[1]{>{\raggedright\let\newline\\\arraybackslash\hspace{0pt}}m{#1}}

\newif\ifpreprint

\newcommand{\preprintfooter}{
  \ifpreprint
    \renewcommand{\headrulewidth}{0pt}
    \renewcommand{\footrulewidth}{0pt}
    
    \fancypagestyle{plain}{
      \fancyhf{}
      \fancyfoot[C]{\textcolor{red}{\bf {\large This is a preprint under submission. Please do not distribute.}}}
    }
    \pagestyle{plain}  
  \fi
}

\begin{document}

\date{}

\title{\Large \bf \textsc{Memory DisOrder}: Memory Re-orderings as a Timerless Side-channel}



\newcolumntype{Y}{>{\centering\arraybackslash}X}


\author{
    \noindent\begin{tabularx}{\linewidth}{Y Y Y}
        {\rm Sean Siddens}$^{\dagger}$ & {\rm Sanya Srivastava} & {\rm Reese Levine} \\
        University of Washington & Duke University & UC Santa Cruz \\
    \end{tabularx}
    \\ \\
    \begin{tabularx}{\linewidth}{Y Y}
        {\rm Josiah Dykstra}$^{\dagger}$ & {\rm Tyler Sorensen}$^{\dagger}$ \\
        Raytheon BBN Technologies & Microsoft Research \\ & UC Santa Cruz \\
    \end{tabularx}
}


\maketitle

\renewcommand{\thefootnote}{\fnsymbol{footnote}}
\footnotetext[2]{Work performed while at Trail of Bits.}
\renewcommand{\thefootnote}{\arabic{footnote}}

\preprintfooter

\begin{abstract}
To improve efficiency, nearly all parallel processing units (CPUs and GPUs) implement relaxed memory models in which memory operations may be re-ordered, i.e., executed out-of-order. Prior testing work in this area found that memory re-orderings are observed more frequently when other cores are active, e.g., stressing the memory system, which likely triggers aggressive hardware optimizations. 

In this work, we present \papername: a  timerless side-channel that uses memory re-orderings to infer activity on other processes. We first perform a fuzzing campaign and show that many mainstream processors (X86/Arm/Apple CPUs, NVIDIA/AMD/Apple GPUs) are susceptible to cross-process signals. We then show how the vulnerability can be used to implement classic attacks, including a covert channel, achieving up to 16 bits/second with 95\% accuracy on an Apple M3 GPU, and application fingerprinting, achieving reliable closed-world DNN architecture fingerprinting on several CPUs and an Apple M3 GPU. 
Finally, we explore how low-level system details can be exploited to increase re-orderings, showing the potential for a covert channel to achieve nearly 30K bits/second on X86 CPUs. 
More precise attacks can likely be developed as the vulnerability becomes better understood.


\end{abstract}

\section{Introduction}

Modern shared memory parallel processors implement complex pipelines and memory systems. This complexity increases efficiency but has also led to security issues, especially in multiprocessing systems, where sensitive information from one process can be obtained by another (potentially malicious) process. Side-channel attacks are especially nefarious, as the attacker can utilize low-level device details to detect subtle properties of an unassuming victim process.

There are many types of side-channels, utilizing many different system components. Some side-channels require physical access~\cite{xiang2020open}, while others require access to low-level APIs, e.g., to monitor energy~\cite{gao2024deep}. The attack requirements are referred to as the \textit{capability} of the attack. A \textit{low capability} attack has few requirements and, thus, can be deployed in more situations. For example, classic cache-based side-channels require precise timers~\cite{liu2015last-level}. However, the capability of the attack can be lowered by removing the timers and instead using other (seemingly innocent) mechanisms~\cite{schwarz2023fantastic}.


\paragraph{\papername} In this work, we present \papername (or simply \papernameshort): a low-capability side-channel for parallel processors. \papernameshort utilizes memory re-orderings that arise due to hardware relaxed \textit{memory consistency models} (or MCMs)~\cite{nagarajan2020primer}. That is, to increase efficiency, mainstream consumer parallel processors allow memory operations to execute out-of-order, as specified by their MCM. These behaviors occur both on CPUs~\cite{10.1145/1785414.1785443, alglave2009arm} and GPUs~\cite{levine2023gpuharbor}. 
Memory re-orderings can be observed in small parallel programs, often called \textit{litmus tests}, which typically execute with two threads with two memory operations each. By inspecting certain variable and memory values after the test execution, it can be determined if a re-ordering occurred.

Prior works have run these litmus tests across many different processors to test conformance to an MCM specification. When testing GPUs, it was found that the number of observed re-orderings could be amplified by utilizing additional threads that added stress to the system~\cite{asplos15}. Inspired by these results, this paper explores the potential of memory re-orderings as a side-channel, where one (potentially malicious) process repeatedly executes a litmus test, requiring only the ability to launch threads and execute memory loads and stores. 
Other (victim) processes may create identifiable patterns of system pressure when executing certain programs, causing re-orderings to be observed by another process. 

\begin{figure}[t]
\begin{tabularx}{\columnwidth}{X X}
         \multicolumn{2}{c}{Initially: \texttt{*x == 0 \&\& *y == 0}} \\
    \midrule
    \underline{thread 0} & \underline{thread 1} \\
    \circledBlack{\vphantom{A}a}\hspace{.1cm} \texttt{Write(x, 1)} &  
\circledWhite{\vphantom{A}c}\hspace{.1cm} \texttt{v0 = Read(y)}  \\
        \circledBlack{\vphantom{A}b}\hspace{.1cm} \texttt{Write(y, 1)} &
        \circledWhite{\vphantom{A}d}\hspace{.1cm} \texttt{v1 = Read(x)}  \\
        \midrule
        \multicolumn{2}{c}{re-order check: \texttt{v0 == 1 \&\& v1 == 0}}
    \end{tabularx} \hspace{1cm}
     \caption{The Message Passing (MP) litmus test. Thread 0 writes to \texttt{x} and then \texttt{y}. Thread 1 reads the values in the opposite order. If thread 1 observes the updated value for \texttt{y} before the updated value for \texttt{x}, then a re-ordering occurred.}
     \label{fig:mp}
\end{figure}

\begin{figure}[t]
    \centering
    \includegraphics[width=\columnwidth]{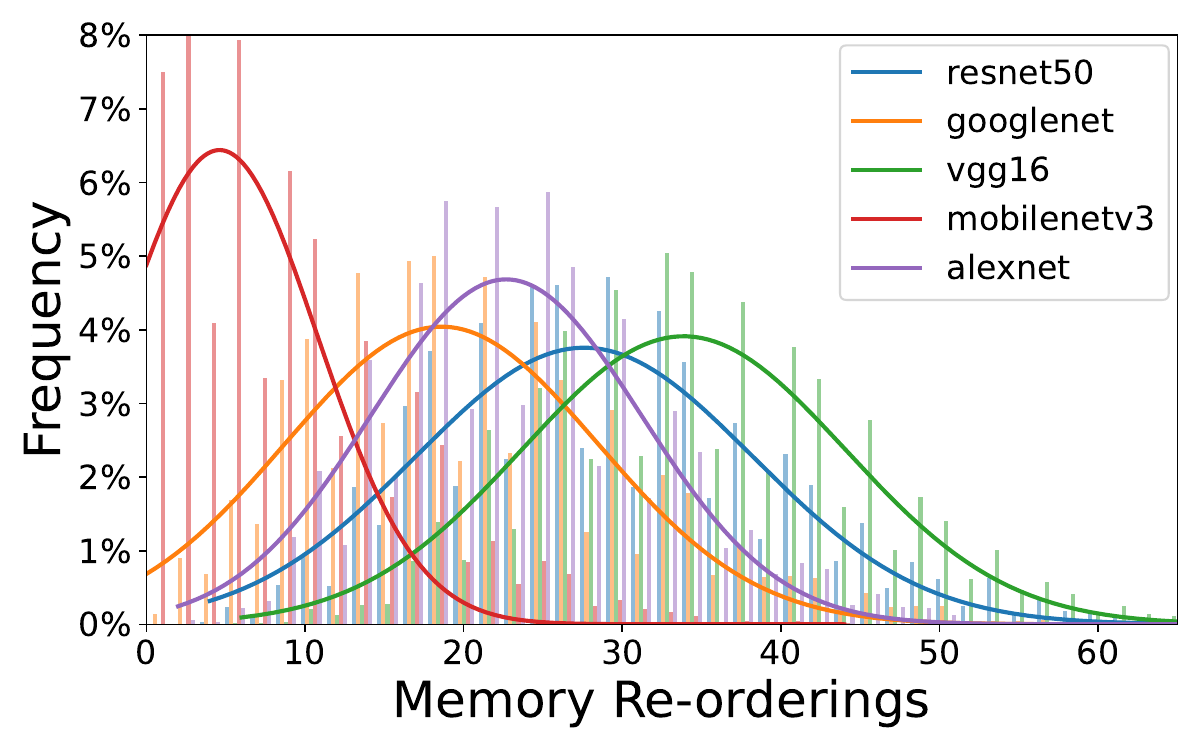}
    \caption{Histogram of re-ordering observations by an attacker when a victim is running inference from common DNN architectures on an Apple M3 GPU.} \label{fig:intro-histogram}
\end{figure}

\subsection{Example: DNN Architecture Fingerprinting \label{sec:fingerprinting-intro}}

We illustrate the potential of \papernameshort using an example in which there are two processes: a victim and an attacker. The victim process repeatedly runs DNN inference on an Apple M3 GPU, which is known to have a relaxed MCM~\cite{metal-shading-language}. The attacker repeatedly executes a classic litmus test, known as Message Passing (or MP), shown in \myfig\ref{fig:mp}. In order to satisfy the re-order check for this test, either the writes on thread 0 or the reads on thread 1 must be re-ordered. Because these tests are short, the attacker can execute many iterations while the victim is executing DNN inference. 

\myfiglong\ref{fig:intro-histogram} shows a histogram of the frequency of observed memory re-orderings by the attacker when the
victim is running 5 different (common) DNN architectures: resnet50, googlenet, vgg16, mobilenetv3\_small (reported as mobilenetv3 in the figure), and alexnet. 
We see that each DNN produces a unique signature, i.e., an approximate normal distribution of observed re-orderings. We then split these observations into a training and testing set (discussed in \mysec\ref{sec:fingerprinting}) and show that an attacker can automatically classify which DNN architecture a victim is running.

\subsection{\papername Overview}
This paper provides a detailed investigation into the potential of memory re-orderings as a side-channel. While there are many low capability side-channels, e.g., see ~\cite{yu2023synchronization}, we believe \papername has a range of unique properties:

\begin{enumerate}
    \item \textit{Low capability}: \papernameshort  only requires two threads and shared memory; it does not require timers, performance counters, or physical access. 
    \item \textit{Wide reach}: Mainstream processors (CPUs and GPUs) implement relaxed MCMs, which are accessible through common programming languages, e.g., C++~\cite{batty_c11}.
    \item \textit{Empirically based}:  Many side-channels target low-level hardware components through meticulous reverse engineering~\cite{ahn2021network, naghibijouybari2017constructing, yu2023synchronization}. \papernameshort attacks can be constructed using simple empirical fuzzing techniques. 
    \item \textit{Unknown potential and mitigations}: 
    The exact hardware component(s) that cause a memory re-ordering are not documented or well-understood. As such, it is difficult to develop rigorous mitigations, and more targeted attacks can likely be developed through detailed explorations.
\end{enumerate}



This work focuses on the \textit{feasibility} and \textit{breadth} of \papernameshort: 
we show that six different architectures are impacted, spanning both CPUs as well as GPUs; furthermore, we show that this vulnerability crosses virtualization boundaries (KVM on Linux). While it is not feasible to reverse engineer the complex hardware interactions that cause memory re-orderings across all these devices, we provide an initial investigation into how more targeted approaches can improve re-ordering observations (i.e., signals) by several orders of magnitude on X86 CPUs (\mysec\ref{sec:arch-aware}). In other cases, we provide hypotheses and reference documentation when available and leave more detailed explorations to future work. Thus, \papernameshort attacks can be developed with either (1) low precision using only simple empirical data, or (2) higher-precision, constructed by  thorough reverse engineering. 


We found only one prior security work proposing re-ordering side-channels~\cite{sophia-out-of-order}. However, they only run small simple experiments on X86 CPUs and do not discuss how to search for signals systematically across different architectures. 
Additionally, \papernameshort attacks require two threads and shared memory; it has been shown that these ingredients are often sufficient to construct a high-resolution timer (e.g., see~\cite{schwarz2023fantastic}), and thus, re-enable classic cache-based side-channels. However, given that the exact reasons for memory re-orderings are still largely unexplored, we believe there is novel potential in \papernameshort and it should \textit{not} be considered only a proxy for cache-based side-channels.

\begin{table*}[]
    \centering

        \caption{The devices and configurations used in our study. We observe \papernameshort on all of our devices. We construct additional attacks for Arm, X86, M1-CPU and M3-GPU (highlighted in green).     \label{tab:devices-tested}}

\vspace{.1cm}
        
    \begin{tabularx}{\textwidth}{l l l l l X}
    \toprule
         \textbf{Type} & \textbf{Vendor} & \textbf{Name} & \textbf{Short Name} & \textbf{Cores or CUs} & \textbf{Configurations}  \\
         \hline
        \rowcolor{green!20} CPU & Apple & M1 & M1-CPU & 8 & MacOS different processes\\
        \rowcolor{green!20} CPU & Arm & A78 (Nvidia Jetson Orin Nano) & Arm & 6 & Linux different processes, KVM\\
        \rowcolor{green!20} CPU & Intel & i7-12700K & X86 & 12 & Linux different processes, KVM\\
        \hline
        \rowcolor{green!20} GPU & Apple & M3 & M3-GPU & 10 & MacOS different processes\\
        GPU & NVIDIA & GeForce RTX 4070 & NVIDIA & 46 & Linux different processes\\
        GPU & AMD & Radeon RX 7900 XT & AMD & 84 & Linux different processes \\
         \hline
    \end{tabularx}

\end{table*}

\paragraph{Fuzzing for \papernameshort }
We begin by designing a fuzz testing campaign to determine if it is possible to observe a signal across processes using memory re-orderings (\mysec\ref{sec:fuzz-testing}). The fuzz testing contains two parameterized processes: the \textit{Stressor} and the \textit{Listener}. 
The Listener runs memory re-ordering litmus tests, while the Stressor applies system stress. We record the number of re-ordering observations both with and without the Stressor executing and check if there is a difference. If so, it is possible for the Listener to determine if it is running concurrently with the Stressor.  


We explore six devices: three CPUs, and three GPUs, listed in \mytab\ref{tab:devices-tested}. 
Although the strength varies, we were able to observe signals across these devices and configurations. 
We say that devices for which we can observe a signal are vulnerable to \papernameshort.

\paragraph{\papernameshort Attacks}
We then select a subset of our devices (highlighted with green in \mytab\ref{tab:devices-tested}) and utilize data from our fuzzing campaign to implement two classic attacks: covert channels (\mysec\ref{sec:covert-channels}) and application fingerprinting (\mysec\ref{sec:fingerprinting}). For the Arm and X86 devices, we carry out these attacks across KVM virtualization boundaries.

To implement a covert channel, we identify test and stress combinations that can provide a high and low signal, and use that to implement a covert channel that can communicate reliably at up to 16 bits/second (on an Apple M3 GPU).
%
To implement application fingerprinting, we run a selected listener alongside the victim application. The collected data is split into a training and test set to simulate an attack. We show that a listener (attacker) needs roughly only 5 seconds to successfully classify which candidate DNN architecture is running with up to 100\% accuracy on an Intel X86 CPU. We also show that other application activity, e.g., launching Google Chrome, provides significant \papernameshort signals.

\myseclong\ref{sec:arch-aware} concludes with an exploration on how low-level system details can be exploited to increase re-orderings, and thus, increase attack precision. We show that on X86 CPUs exercising certain L1 cache sets dramatically increases the number of re-orderings, with the potential to increase the covert channel rate to nearly 30k bits/second. This exploration can serve as the foundation for constructing more precise \papernameshort attacks. 

\paragraph{Contributions}
In summary, our contributions are:
\begin{itemize}
    \item \papername: A novel low capability side-channel attack that utilizes memory re-orderings.
    \item A fuzz testing campaign that shows many mainstream CPUs and GPUs are vulnerable to \papernameshort (\mysec\ref{sec:fuzz-testing}). 
    \item Illustrating how to construct \papernameshort covert channels (\mysec\ref{sec:covert-channels}) and application fingerprinting (\mysec\ref{sec:fingerprinting}).
     \item An exploration of how low-level system details can be exploited to dramatically increase re-orderings used in \papernameshort  (\mysec\ref{sec:arch-aware}).
\end{itemize}

\noindent
Mitigations and remediations are discussed in \mysec\ref{sec:mitigations} and responsible disclosure is discussed in \mysec\ref{sec:ethics}.

\section{Background} \label{sec:background}

Because we target both CPUs and GPUs, we give a brief background on their multiprocessing capabilities and define common terminologies. Next, we provide an overview of memory consistency models and the re-orderings they allow. 

\subsection{Parallel Architectures and Multiprocessing \label{sec:background-devices}}

Nearly all mainstream computing devices (laptops, phones, servers) contain parallel processing units (e.g., CPUs, GPUs) that allow hardware resources to execute tasks simultaneously. These processors typically allow multiprocessing, where tasks from different processes can execute concurrently. Processes provide a clear security boundary---each process should be isolated from all other processes. Virtualization provides an additional security boundary in that malicious processes are not able to utilize operating system (OS) vulnerabilities.

\paragraph{CPUs} CPUs are the processing center of a system. They execute most of the low-level system logic (e.g., the OS), and are the main computational component for many interactive applications, such as web browsers.
CPUs are latency optimized, where a small number of complex \textit{cores} optimize a single stream of instructions (commonly called a \textit{thread}). Different threads are often mapped to different cores. In some systems, such as Linux, software threads can be pinned to cores using a low-level API. Mainstream CPU operating systems (Linux, MacOS) provide parallel multiprocessing where different processes can execute in parallel on different cores.


\paragraph{GPUs} GPUs are highly parallel programmable accelerators. Their parallelism is organized hierarchically, where the base unit is a \textit{processing element} which computes a stream of computation (again, called a thread). Processing elements are organized into \textit{compute units} (CUs)\footnote{Called \textit{streaming multiprocessors} (SMs) in NVIDIA documentation; however, we use the more portable Khronos terminology in this work}. As opposed to CPUs, GPUs are throughput oriented, computing many tasks in parallel using simpler hardware components per task.

A GPU program is called a \textit{kernel}; it is launched from a host (CPU) program that specifies the hierarchical configuration of threads, namely the total number of threads and a partition of the threads into \textit{workgroups}, which are guaranteed to execute on the same CU. While GPUs have a complex execution model, e.g., where some threads are executed synchronously, these details are not required for \papernameshort. 

GPU multiprocessing capabilities and properties vary across systems and configurations. Classically, GPUs do \textit{not} execute kernels from different processes in parallel; instead, they execute different kernels sequentially~\cite{sorensen2024leftoverlocalslisteningllmresponses}. However, it is becoming more common for GPUs to support \textit{parallel kernel execution}, where different kernels are executed on different compute units e.g., through configurations like NVIDIA's MPS~\cite{mps} and MIG~\cite{mig}. In \mysec\ref{sec:gpu-results} we empirically show that Apple GPUs also provide parallel kernel execution.

\paragraph{CPU/GPU Commonalities}  Both CPUs and GPUs provide a general programming interface that allows many threads to be launched and access memory without elevated privileges. These memory accesses reliably get translated to hardware memory accesses, i.e., without compiler optimizations, using \texttt{atomic} libraries.
Furthermore, both CPUs and GPUs contain complex memory hierarchies, which contain at least two levels of caches and a main memory. This common design is likely somewhat responsible for observable memory re-orderings on both devices.

\subsection{Memory Consistency Models}

A memory consistency model (MCM) defines the values that load instructions are allowed to return in a shared memory parallel program. There are both language models (e.g., for C++~\cite{batty_c11}) and hardware models (e.g., for X86~\cite{x86-formal} and Arm~\cite{10.1145/2837614.2837615}). These models can be complex and have been developed and refined over many years.
However, this work does not require the full complexity of state-of-the-art MCMs. Our initial exploration of \papernameshort only requires a simple foundation for empirical testing, which we describe below.

\paragraph{Instruction Re-ordering MCMs} This work considers MCMs that can be completely defined in terms of thread-local re-orderings, a subset of the models described in~\cite{arvind}, which we will call IR (Instruction Re-ordering) models. For example, an IR model might allow a \textit{Write-Read} (or \textit{WR}) re-ordering. That is, if a program contains a write ($W$) followed by a read ($R$), then they are allowed to be re-ordered. Given that we only consider two memory instructions, write ($W$) and read ($R$), there are only four possible pairs that are can be re-ordered (\textit{RR}, \textit{RW}, \textit{WR}, \textit{WW}). An IR model can be defined by enumerating which of these re-orderings are allowed. 

IR models typically do not allow re-ordering of accesses that target the same location, as this could break single-threaded sequential execution. That is, an IR model that allows \textit{RW} re-orderings will only allow the re-ordering if the two accesses target different locations. When we discuss each of our test processors (\mysec\ref{sec:fuzz-testing}), we will discuss which IR MCM most closely approximates the allowed behaviors for that processor.

\begin{table}
    \centering
    \small
        \caption{Generalizing the MP test of \myfig\ref{fig:mp} creates five additional classic litmus tests.}

\vspace{.1cm}

    \begin{tabular}{l l l l l l l}
    \toprule
        \textbf{Test Name} & \circledBlack{a} & \circledBlack{b} & \circledWhite{c} & \circledWhite{d} & \textbf{re-order check}\\
        \midrule
        Message Passing (MP) & $W$ & $W$ & $R$ & $R$ & \texttt{v0=1 \& v1=0} \\
        Store Buffering (SB) & $W$ & $R$ & $W$ & $R$ & \texttt{v0=0 \& v1=0}\\
        Load Buffering (LB) & $R$ & $W$ & $R$ & $W$ & \texttt{v0=1 \& v1=1}\\
        2+2W & $W$ & $W$ & $W$ & $W$ & \texttt{*x=1 \& *y=2} \\
        Store (S) & $W$ & $W$ & $R$ & $W$ & \texttt{*x=2 \& v0=1} \\        
        Read (R) & $W$ & $W$ & $W$ & $R$ & \texttt{*y=2 \& v0=0} \\
        \bottomrule
    \end{tabular}
    \label{tab:all-tests}
\end{table}

\paragraph{IR Litmus Tests} The simplicity of IR models allows them to be thoroughly tested using simple litmus tests. We can generalize the MP test of \myfig\ref{fig:mp} where each instruction (\circledBlack{a}, \circledBlack{b}, \circledWhite{c}, \circledWhite{d}) can be instantiated with an \textit{R} or a \textit{W}. There are always two memory locations, \texttt{x} and \texttt{y}. Memory operations \circledBlack{a} and \circledWhite{d} target \texttt{x} and memory operations \circledBlack{b} and \circledWhite{c} target \texttt{y}. Write operations store unique values (starting at 1 and incrementing with each write), and read operations store to unique variables (starting at \texttt{v0} and incrementing with each load). Each test can check for up to two re-orderings, one per thread. For example, the MP test of \myfig\ref{fig:mp} tests for both a \textit{WW} re-ordering (thread 0) and \textit{RR} re-ordering (thread 1).

Following this formula, we can construct 6 litmus tests, where a post condition can be used to check for IR re-orderings. These tests correspond to classic litmus tests used throughout MCM literature, with admittedly cryptic names. We describe these tests in \mytab\ref{tab:all-tests}, and they can be viewed more concretely in prior works, e.g., ~\cite{kirkham2020foundations}. Re-orderings are not unique for each test, e.g., both SB and R test for the \textit{WR} re-ordering.

\begin{table}[]
        \centering
    \caption{The different testing frameworks (shaded in green), and stresses (shaded in blue), the device they target (CPU or GPU), and the number of fuzzed parameters.    \label{tab:fuzzed-summary}}

    \vspace{.1cm}

    \begin{tabular}{l l r}
    \toprule
    \textbf{Name} & \textbf{Device} & \textbf{Fuzzed Parameters} \\

     \hline
       \rowcolor{green!20} Basic  & CPU & 2\\
        \rowcolor{green!20}Litmus7~\cite{10.1007/978-3-642-19835-9_5}& CPU & 0\\
       \rowcolor{green!20} Perpetual~\cite{melissaris2020perple} & CPU & 2\\
        \rowcolor{green!20} GPU Parallel~\cite{levine2023mc} & GPU & 4\\
     \rowcolor{blue!20}Memory~\cite{kirkham2020foundations} & CPU, GPU & 6 \\
    \rowcolor{blue!20} Thread Launch & CPU & 2 \\
     \hline
    \end{tabular}
    
    \vspace{.2cm}

\end{table}

\section{Fuzz Testing for \papernameshort Vulnerabilities \label{sec:fuzz-testing}}

We now detail a fuzz testing methodology that can check if devices are vulnerable to \papernameshort. 
The methodology consists of two processes: the \textit{Listener}, and the \textit{Stressor}. 
The Listener can be instantiated using one of four different memory ordering testing techniques and the Stressor can be instantiated with one of two system stress techniques. 
For each technique (testing and stressing), we summarize the approach and list the fuzzed parameters, summarized in \mytab\ref{tab:fuzzed-summary}. 

\subsection{The Listener}

The Listener process runs litmus tests for many \textit{iterations} (typically over 100K) and records the number of re-orderings. Although litmus tests are simple, running the tests can be complex, utilizing heuristics to increase throughput and re-orderings.
A \textit{testing framework} takes a litmus test and some tunable parameters. 
We use three frameworks from prior works and provide one new framework.

\paragraph{Basic Testing Framework}
We provide the \textit{basic testing framework} new to this work. While it is simple, it provides some of the most reliable signals when combined with the right stress (see the M1-CPU results in \mysec\ref{sec:cpu-results}). This framework implements the litmus test using C++ threads and relaxed atomic memory accesses, which allows all IR memory re-orderings. For each test iteration, the threads are launched, then joined, then the re-ordering condition is checked. 

This framework contains two fuzzing parameters: the indices of the two memory locations, i.e., \texttt{x} and \texttt{y}. 
Fuzzing the indices allows them to be located across (or within) different memory regions, such as cache lines or memory pages, which could encourage different types of re-orderings to appear. We only implement this framework for CPU systems, as GPUs do not widely support C++ concurrency constructs.

\paragraph{Litmus7 Framework}
The Litmus7 tool~\cite{10.1007/978-3-642-19835-9_5} implements several heuristics on top of the basic framework. For example: memory operations are implemented using inline assembly and threads are not relaunched each iteration, instead using inter-thread synchronization to perform the re-order check and align the next iteration. 
%
Furthermore, Litmus7 performs its own fuzzing, randomly assigning memory locations and permuting how software threads map to test threads. Given this, we run this framework with its default settings and do not fuzz any of our own parameters. Because Litmus7 only targets CPUs (previous GPU versions~\cite{asplos15} are not publicly available), we only test CPU systems with this framework.

\paragraph{Perpetual Framework}
Perpetual testing was used in early MCM testing~\cite{archtest}, and has seen a recent resurgence~\cite{ melissaris2020perple, SrivastavaSanya2024TMMo}. This approach increases throughput by eliminating synchronization by launching testing threads only once. Write operations store algebraic sequences, while the read operations maintain a log of observed values. The log is analyzed at the end of the run; traces that satisfy certain algebraic constraints indicate that a memory re-ordering  occurred. This approach is attractive for \papernameshort as it provides fine-grained observations, potentially enabling more precise attacks.

In order to enable wider testing, we re-implement the X86-exclusive approach of~\cite{melissaris2020perple} using C++ atomic operations and threads. The fuzzed parameters for this approach are the same as for the basic framework, i.e., two memory indices.  Additionally, without significant modifications, this approach requires two read operations in the test; thus, we limit the litmus tests for this testing framework to MP, SB, and LB. In \mysec\ref{sec:fuzzing-results}, we find that unfortunately, this framework does not yield any re-ordering observations on Arm or X86 CPUs. It may be the case that our implementation needs to be more finely tuned, as prior work~\cite{melissaris2020perple} was able to observe X86 re-orderings. 
Because this framework works best executing a small number of threads rapidly, it is more suited for CPUs, and, thus, we do not provide a GPU implementation.



\paragraph{GPU Parallel Testing Framework} This framework~\cite{levine2023mc} utilizes GPU parallelism to enable high testing throughput. Given $N$ GPU threads, this framework instantiates $N$ litmus test to execute in parallel. Each thread executes two instances of a test: acting first as thread 0, and next as thread 1. 
This approach is implemented in WebGPU~\cite{w3c_webgpu}, making it portable across many GPUs. There are four fuzzed parameters in this framework: the number of workgroups, workgroup size, padding between memory locations, and how frequently to synchronize test threads. Litmus test threads are also randomly mapped to GPU threads each iteration.

\subsection{The Stressor}

The Stressor executes a \textit{stress}: a program designed to stress the system and increase memory re-orderings observed by the Listener.  In this work, we explore two stress techniques that we found increased memory re-orderings. One technique is inspired by prior work, while the other is new to this work. These stress processes execute indefinitely, until killed. 

\paragraph{Memory Stress} As the name suggests, this stress targets the memory system and closely follows prior work in GPU MCM conformance testing~\cite{kirkham2020foundations}. This stress allocates a memory buffer and partitions it into stress lines (conceptually similar to cache lines). It then launches threads and maps them to an initial stress line. The threads repeatedly access a stress line with a pattern of loads and stores. After some iterations, the threads move to another stress line.

We provide C++ and WebGPU implementations of memory stress, applying memory stress to CPU memory model testing for the first time. This technique has 6 fuzzed parameters: the stress line size, a thread offset into the stress line, how many iterations to target a given stress line, the stride used to update the memory location, the memory access pattern (a sequence of loads and stores), and the number of stressing threads.

\paragraph{Thread Launch Stress} This stress was discovered serendipitously as we were running early experiments. We found that opening a new terminal while a test was running significantly increased the number of  re-orderings. We distilled this behavior into a stress that repeatedly launches and joins threads. 

We implement this stress in C++. Threads execute a loop containing relaxed atomic memory accesses so that the compiler doesn't optimize away the loop. The two fuzzed parameters for this stress are the number of threads and the thread loop iterations, which controls the rate at which the threads are re-launched. Given that GPUs don't have a similar mechanism to launch threads, instead requiring a whole kernel to be launched, we only provide this stress for the CPU.

\subsection{Checking for \papername}

To check for \papernameshort, the Listener is instantiated with a testing framework and its inputs, i.e., a litmus test and any required parameters. It then executes the litmus test for a given number of \textit{litmus test iterations}. This is called a \textit{test run}. After a test run, the Listener reports on how many re-orderings were observed, i.e., the \textit{re-ordering frequency}.
The Stressor is instantiated with a stress along with any required parameters. We say that the Stressor is able to \textit{signal} the Listener if the re-ordering frequency observed in the Listener is reliably higher when the Stressor is executing. If a signal is observed, then we say that the system is vulnerable to \papernameshort. 

To check for a reliable signal, the Listener performs $X$ test runs in isolation, creating a set $B$ (baseline) with $X$ samples of re-ordering frequencies. Then, the Stressor begins, running indefinitely until killed. The Listener performs $X$ test runs again, executing simultaneously with the Stressor, creating another set $S$ (stressed) of samples. These sets are then compared with a statistical method (Mann-Whitney U test~\cite{mann1947test}) to determine if values in one set are larger than another. 
We also compute the Common Language Effect Size (CLES), which is the probability that a randomly selected element from $S$ is larger than a randomly selected element from $B$. If the CLES is 100\%, i.e., all elements in $S$ are larger than $B$, then we say that the signal is \textit{reliable}. Furthermore, for each test in which a signal is observed, we record the average percent increase in observed memory re-orderings. Each of these statistical tests, i.e., consisting of $X$ test runs, both with and without stress, is called a \textit{trial}.

\paragraph{Fuzzing}
To fuzz for \papernameshort, a simple script can be run that iterates through each stress $s$, each testing framework $f$ (\mytab\ref{tab:fuzzed-summary}), and each applicable litmus test $t$ (\mytab\ref{tab:all-tests}).  The script instantiates the Listener with $f$, and uses $t$ as the input test. The Stressor is instantiated with $s$. The fuzzer then randomly selects the additional parameters and performs a trial. The fuzzer executes many signal tests and records their outcomes. 

\section{Fuzzing Results \label{sec:fuzzing-results}}

We now detail the results of running the fuzz testing campaign across the devices in \mytab\ref{tab:devices-tested}, all of which we are able to see statistically significant (and reliable) signals. \textit{Thus, \papernameshort vulnerabilities exist widely on mainstream CPUs and GPUs}. We select 10 as the number of test runs to create the sample sets ($B$ and $S$) that are compared in the trial. We configure the fuzzing campaign for each device such that it finishes executing roughly in 8 hours (overnight); this includes setting parameters like litmus test iterations and the number of trials. 

We summarize our findings across all processors and configurations in \mytab\ref{tab:all-results}. For each configuration, we report on the percent of trials that showed a signal, both \textit{any} signal, i.e., there was a statistical difference in the observation sets, and a \textit{reliable} signal, i.e., the CLES was 100\%. Note that the reliable percent will be less than (or equal to) the any signal percent. We summarize the percent increase in re-orderings observed, both the average and the max. We then show the average CLES across all the signals to understand how frequently increases in re-orderings were observed. Finally, we show how many trials were run. The number of trials are different across devices and configuration given the different speed of the processors and throughput of the testing framework. 

On Arm and X86 CPUs running Linux, we are able to explicitly set thread affinities, i.e., map program threads to hardware cores. 
For these devices, we consider two configurations related to the attacker capability: one where the affinities can be explicitly mapped (called \textit{explicit pinning}) and one where affinities are managed by the OS. Explicit pinning is strictly a higher capability than OS-managed, as it requires an attacker to be able to explicitly map the Listener and Stressor to specific device cores. To find an affinity mapping, we ran several pilot experiments across a range of mappings and selected the one that had the highest average CLES.

Given that each device contains its own quirks and considerations, we discuss them separately below.

\subsection{CPU results \label{sec:cpu-results}}

\paragraph{Arm} The Arm processor we test has 6 cores and runs Ubuntu 22.04.
Arm devices have a famously relaxed memory model~\cite{10.1145/2837614.2837615}, 
but the specification tends to be more permissive than what is implemented in practice.
The LB test is not observable on any Arm system, thus we omit it from the fuzzing campaign for this processor. For the explicit affinity mapping, we fixed the test threads to only run on cores 0 and 5, while the stress threads were fixed to cores 1 through 4.

Thread launch stress coupled with the basic testing framework provides the most reliable signals for this device. 
Reliable signals can be obtained roughly 12\% or 64\% of the time, depending on whether the affinity is explicitly set, highlighting the increased benefit of this capability. In both cases, the maximum increase in re-orderings observed is in the order of 1K percent, providing easily distinguishable signals. Memory stress provides signals less frequently, at most in 25\% of tests and reliable signals at most in 7\% using the litmus7 testing framework \textit{without explicit} pinning; however, the maximum increase in re-orderings is similar to thread launch stress. Despite the potential for increased precision of perpetual tests, we were unable to observe any signals using this method.



\begin{table*}[]
\centering
\caption{Results of the fuzz testing campaign across our devices. \label{tab:all-results}}

\vspace{.1cm}
          
\begin{tabular}{l l l l r r r r r r}
\toprule
                      &  &  &   & \multicolumn{2}{c}{\textbf{Signal \%}} &  \multicolumn{2}{c}{\textbf{Increase \%}} \\
\cmidrule(lr){5-6}\cmidrule(lr){7-8}
\textbf{Device}       & \textbf{Pinning?} & \textbf{Testing F.} &  \textbf{Stress} & \textbf{Any} & \textbf{Reliable}  &  \textbf{Avg.} & \textbf{Max} &  \textbf{CLES (Avg.)} & \textbf{Trials} \\
      
\hline
      
\multirow{12}{*}{Arm} & \cellcolor{red!20}                           & \cellcolor{blue!20}                               & memory        & 10.93 &  0.94 & 519 & 1703  & 0.86 & 320 \\         
                      & \cellcolor{red!20}                           & \cellcolor{blue!20}   \multirow{-2}{*}{basic}     & thread launch & 59.06 & 11.87 & 432 & 2700  & 0.92 & 320 \\
                      & \cellcolor{red!20}                           & \cellcolor{yellow!20}                             & memory        & 24.69 &  7.19 & 132 & 411   & 0.92 & 320 \\
                      & \cellcolor{red!20}                           & \cellcolor{yellow!20} \multirow{-2}{*}{litmus7}   & thread launch & 36.87 &  0.94 & 77  & 205   & 0.88 & 320 \\
                      & \cellcolor{red!20}                           & \cellcolor{orange!20}                             & memory        &  0.00 &  0.00 & 0   & 0     & 0.00 & 256 \\
                      & \cellcolor{red!20}\multirow{-6}{*}{no} & \cellcolor{orange!20} \multirow{-2}{*}{perpetual} & thread launch &  0.00 &  0.00 & 0   & 0     & 0.00 & 256 \\

                      & \cellcolor{green!20}                          & \cellcolor{blue!20}                               & memory        & 17.92 &  4.16 & 424 & 2590 & 0.93 & 240 \\
                      & \cellcolor{green!20}                          & \cellcolor{blue!20}   \multirow{-2}{*}{basic}     & thread launch & 85.00 & 64.16 & 296 & 1097 & 0.97 & 240 \\
                      & \cellcolor{green!20}                          & \cellcolor{yellow!20}                             & memory        &  0.83 &  0.41 &  27 &  38 & 0.90 & 240 \\
                      & \cellcolor{green!20}                          & \cellcolor{yellow!20} \multirow{-2}{*}{litmus7}   & thread launch & 22.92 &  0.42 &  33 &  106 & 0.86 & 240 \\
                      & \cellcolor{green!20}                          & \cellcolor{orange!20}                             & memory        &  0.00 &  0.00 &   0 &    0 & 0.00 & 128 \\
                      & \cellcolor{green!20}\multirow{-6}{*}{yes} & \cellcolor{orange!20} \multirow{-2}{*}{perpetual} & thread launch &  0.00 &  0.00 &   0 &    0 & 0.00 & 128 \\

\hline

\multirow{12}{*}{X86} & \cellcolor{red!20}                           & \cellcolor{blue!20}                               & memory        &   3.12 &  0.00 & 109 & 170 & 0.84 & 128 \\
                      & \cellcolor{red!20}                           & \cellcolor{blue!20} \multirow{-2}{*}{basic}       & thread launch &   0.00 &  0.00 &   0 &   0 & 0.00 & 128 \\
                      & \cellcolor{red!20}                           & \cellcolor{yellow!20}                             & memory        &  10.54 &  1.17 &  89 & 365 & 0.89 & 256 \\
                      & \cellcolor{red!20}                           & \cellcolor{yellow!20} \multirow{-2}{*}{litmus7}   & thread launch &  99.61 & 52.73 &  72 & 136 & 0.97 & 256 \\
                      & \cellcolor{red!20}                           & \cellcolor{orange!20}                             & memory        &   0.00 &  0.00 &   0 &   0 & 0.00 & 128 \\
                      & \cellcolor{red!20}\multirow{-6}{*}{no} & \cellcolor{orange!20} \multirow{-2}{*}{perpetual}       & thread launch &   0.00 &  0.00 &   0 &   0 & 0.00 & 128 \\
         
                      & \cellcolor{green!20}                          & \cellcolor{blue!20}                              & memory        &  82.03 &  1.56 &  123 & 274 & 0.88 & 128 \\
                      & \cellcolor{green!20}                          & \multirow{-2}{*}{basic} \cellcolor{blue!20}      & thread launch &  39.84 &  0.00 &   61 & 107 & 0.83 & 128 \\
                      & \cellcolor{green!20}                          & \cellcolor{yellow!20}                            & memory        &   2.34 &  0.00 &   21 &  25 & 0.83 & 256 \\
                      & \cellcolor{green!20}                          & \cellcolor{yellow!20} \multirow{-2}{*}{litmus7}  & thread launch & 100.00 & 71.28 &   50 &  72 & 0.99 & 256 \\
                      & \cellcolor{green!20}                          & \cellcolor{orange!20}                            & memory        &   0.00 &  0.00 &    0 &   0 & 0.00 & 128 \\
                      & \cellcolor{green!20}\multirow{-6}{*}{yes} & \cellcolor{orange!20} \multirow{-2}{*}{perpetual}    & thread launch &   0.00 &  0.00 &    0 &   0 & 0.00 & 128 \\

\hline

\multirow{6}{*}{M1-CPU} & \cellcolor{red!20}                          & \cellcolor{blue!20}                               & memory        & 41.56 & 20.93 &  1910 &  85814 & 0.95 & 320 \\
                        & \cellcolor{red!20}                          & \multirow{-2}{*}{basic} \cellcolor{blue!20}       & thread launch & 83.43 & 67.81 &  8429 & 770700 & 0.98 & 320 \\
                        & \cellcolor{red!20}                          & \cellcolor{yellow!20}                             & memory        & 68.75 & 38.44 & 14725 & 208033 & 0.94 & 320 \\
                        & \cellcolor{red!20}                          & \cellcolor{yellow!20} \multirow{-2}{*}{litmus7}   & thread launch & 95.31 & 77.50 &  5354 & 120350 & 0.99 & 320 \\
                        & \cellcolor{red!20}                          & \cellcolor{orange!20}                             & memory        &  5.07 &  1.17 &   101 &    326 & 0.87 & 256 \\
                        & \cellcolor{red!20}\multirow{-6}{*}{no} & \cellcolor{orange!20} \multirow{-2}{*}{perpetual} & thread launch & 19.53 &  4.69 &    40 &    125 & 0.88 & 256 \\

\hline

\multicolumn{10}{c}{\textbf{GPUs}} \\

\hline 

AMD    & \cellcolor{red!20}no & GPU Parallel & memory & 13.8 &  5.9 & 7890 & 170300   & 0.89 & 708 \\
NVIDIA & \cellcolor{red!20}no & GPU Parallel & memory &  5.6 &  2.8 & 27768 & 499600  & 0.88 & 726 \\
M3-GPU & \cellcolor{red!20}no & GPU Parallel & memory & 52.4 & 20.5 & 92470 & 4921300 & 0.92 & 696 \\
       
\hline                  

\end{tabular}
   
\end{table*}

\paragraph{X86} Our X86 processor is running on Ubuntu 22.04 and
is documented to have 12 cores, with 8 being hyper-threaded ($\times$2) performance cores and 4 being efficiency cores. X86 systems provide a stronger memory model than Arm~\cite{10.1145/1785414.1785443}, only allowing \textit{WR} re-orderings. Thus, we use SB and R in this fuzzing campaign for this device.
%
Similar to Arm, we test with and without explicit pinning on this device. To determine a good pinning, we ran pilot experiments across a range of affinity mappings and found that the highest average CLES occurred when both the test threads are pinned to performance cores 0 and 1. Given that the performance cores are hyperthreaded, these two cores (0 and 1) are hardware threads that execute on the same core. 

Unlike Arm, X86 provides documentation about the hardware mechanisms that cause memory re-orderings. Each processor contains a \textit{store buffer}, which buffers writes before they are flushed to the memory subsystem. Because store buffers are processor-local, our initial hypothesis was that they would not be susceptible to cross-process stress. However, our results, summarized in \mytab\ref{tab:all-results}, show otherwise. When cores are explicitly pinned, we can observe signals up to 100\% of the time, with 71\% being reliable using the litmus7 testing framework and thread launch stress. In fact, this test and stress combination remains very effective even without explicit pinning. However, memory stress is also able to provide many signals with the basic framework but is more sensitive to explicit pinning (82\% vs. 3\%). Similarly to Arm, the perpetual testing framework did not provide any signals for this device.

While store buffers are processor local, these results show they are susceptible to cross-process memory traffic, which seems to influence when a flush is triggered, e.g., to avoid overloading the memory system. Our extra investigation into X86 (\mysec\ref{sec:arch-aware}) further explores this. These results illustrate just how complicated 
low-level components are, and, thus, how potentially subtle \papernameshort attacks might be. 

\paragraph{M1-CPU} 
The Apple M1-CPU has a total of 8 cores consisting of 4 performance cores and 4 efficiency cores. Although documentation for this processor is sparse, it implements the Arm ISA, and thus allows similar re-orderings. A pilot study revealed that we can observe all IR re-orderings on this device, except for LB (just like Arm). Because of this, we similarly omit LB from the fuzzing campaign for this device. Unlike Linux, MacOS does not support thread affinity mappings, and, thus, cannot run experiments with explicit pinning. 

Reviewing the results for this device in \mytab\ref{tab:all-results}, there are several trends similar to Arm: for example, the most reliable signals come from thread launch stress, particularly when combined with litmus7. However, the largest percent increase occurs with thread launch stress and the basic testing framework, achieving the highest across our CPU experiments, with a max of 770K percent increase. Similar to the other CPUs, memory stress also provides relatively common signals and high increase (more so using Litmus7 than the basic framework). Unlike Arm and X86, perpetual tests show some signals on this processor. However, their reliability and percent increases are low compared to other methods. 





\subsection{GPU Results \label{sec:gpu-results}}
All GPUs we tested are documented to provide very relaxed memory models~\cite{levine2023gpuharbor}. We were able to observe memory re-orderings on \textit{all} tests in \mytab\ref{tab:all-tests} on the GPUs in our study. 
The percent increase in re-ordering observations on GPUs is very high; in fact, memory re-orderings are often not observed at all without some kind of stress (see~\cite{asplos15}). 
To avoid infinite percent increases when the baseline is zero, we treat the baseline count as 1.

\paragraph{AMD Radeon RX 7900 XT} 
Our AMD GPU provides the sequential kernel execution model (recall from \mysec\ref{sec:background-devices}). Despite the Listener and Stressor not executing in parallel on the GPU, prior works have shown that memory effects across GPU kernels can still be observed, e.g., in prime-and-probe style attacks~\cite{giner2024webgpucacheattack}. Our results, shown near the bottom of \mytab\ref{tab:all-results}, reveal that \papernameshort can also be observed in this execution model. Signals on this device are rare, however, there are some reliable signals (less than 6\% of the time), and the maximum percent increase is 170K.

\paragraph{NVIDIA GeForce RTX 4070} 
Similar to the AMD, NVIDIA GPUs provide the sequential kernel execution model by default. Our results show that this device is also impacted by \papernameshort, however frequency and reliability of signals are less than AMD (while the percent increase in re-ordering observations ends up being higher than AMD). 


NVIDIA additionally offers \textit{Multi-Process Service} (MPS)~\cite{mps}, which was developed for when kernels do not contain enough parallelism to utilize the entire GPU. MPS allows multiple kernels to execute on the GPU in parallel, but the documentation warns that this mode will reduce security. 
We were able to run some pilot experiments using this configuration and observed much higher \papernameshort metrics, providing evidence that parallel kernel execution environments cause GPUs to be more vulnerable to \papernameshort.

\paragraph{M3-GPU} Similar to the M1-CPU, there is little documentation about this processor. To demystify its execution model, we designed a small microbenchmark in which two distinct processes each execute a small kernel for roughly the same amount of time $k$. 
We spawn both processes simultaneously and observe that the total time needed for both to finish is also $k$. 
Thus, we can conclude that this GPU provides a parallel kernel execution model. Our results, at the bottom of \mytab\ref{tab:all-results}, show that this parallel kernel execution leads to more frequent, reliable, and effective signals (by at least an order of magnitude) than on the other GPU devices. Thus, if GPUs move towards more parallel execution models, \papernameshort vulnerabilities may become more effective. 


\subsection{Virtualization Boundaries}

Our results thus far have simply tested cross-process signals on readily available consumer devices. However, it is important to test other systems for \papernameshort, especially multi-tenant, security critical systems.
While we showed that X86 systems (which are widely deployed in the cloud) are vulnerable to \papernameshort, Arm systems potentially have a larger attack surface, given that they allow many more re-orderings. We note that Arm-based systems are becoming more common in the cloud, such as Amazon's Graviton processors~\cite{graviton} executing on multi-tenant machines through the Nitro hypervisor~\cite{nitro}. Similarly, Google Cloud provides Arm-based Axiom processors~\cite{axiom} and potentially use KVM as the hypervisor~\cite{cloud-kvm}.  

More recently, GPUs now have some virtualization support. In these approaches, the physical GPU hardware is partitioned, and each virtual GPU is given a physical slice of the GPU. 
NVIDIA offers MIG~\cite{mig}, and AMD offers SR-IOV~\cite{SR-IOV}. We do not test those systems currently for two reasons: they require high-end recent GPUs, which we do not have immediate access to, and they require re-writing some of our testing code.


\section{Implementing \papernameshort Attacks}
Utilizing the data from the fuzzing campaign, we can identify litmus tests, testing frameworks, and stress combinations that expose reliable and easily identifiable signals. We show how this can serve as a foundation for implementing two classic attacks: a covert channel and application fingerprinting. 
Given that these attacks require extensive hand-tuning and reliable signals, 
we only implement them on a subset of our devices, namely all of the CPUs and the Apple GPU. 

\paragraph{Security boundaries} In this section, we increase the security boundaries on our devices when possible. For the Apple CPU and GPU, we show how \papernameshort can cross process boundaries. For Arm and X86, we implement these attacks over a KVM virtualization boundary, launching the Listener process on the host OS and the Stressor process on the guest. 
Similar to the fuzzing campaign, we run some pilot experiments to determine effective thread pinnings. 
On Arm, the KVM instance is allocated three VCPUs, pinned to the host cores 0, 2, and 4 while the Listener process runs the litmus tests on host cores 1 and 3. 
On X86, all experiments allocate ten VCPUs to the KVM instance, which are pinned to even-numbered host cores.
For the covert channel experiment we pin the Listener to host cores 1 and 3 while for the DNN fingerprinting experiment we pin them to host cores 1 and 11.

Our initial experiments show that thread launch stress becomes less reliable across KVM boundaries, which is intuitive given the role of the OS in launching threads. However, memory stress remains reliable, and, thus, attacks on Arm and X86 will use memory stress. 

\subsection{\papernameshort Covert Channels \label{sec:covert-channels}}

\begin{table*}[!t]
\centering
\caption{Details for \papernameshort covert channel implementations. For each device, we show the testing framework and the litmus test used. For each signal, we show the stress technique, the average number of re-orderings (Avg.) and the standard deviation~(Std.). The $\emptyset$ does not have a stress technique as it is simply the absence of stress. X86-arch (highlighted in green) is custom designed to low-level X86 system details and is described more in \mysec\ref{sec:arch-aware}. It does not have a $\emptyset$ signals \label{tab:covert-configs}}

\vspace{.1cm}

\begin{tabularx}{\textwidth}{l l l l r r l r r r r X}
  \toprule
  \textbf{Device} & \textbf{Framework} & \textbf{Test} & \textbf{$\uparrow$ Stress} & \textbf{$\uparrow$ Avg.} & \textbf{$\uparrow$ Std.} & \textbf{$\downarrow$ Stress} & \textbf{$\downarrow$ Avg.} & \textbf{$\downarrow$ Std.} & \textbf{$\emptyset$ Avg.} & \textbf{$\emptyset$ Std.}\\
  \midrule
  Arm    & litmus7      & MP & mem & 5859.1 & 1469.9 & mem & 3224.2 & 709.9 & 437.5 & 141.0\\
  M1     & basic        & R  & TL  &  137.7 &  20.3  & TL  &  37.3  &  7.7  &  3.4  &  2.3\\
  X86    & litmus7      & R  & mem &  803.6 &  46.7  & mem &  648.9 &  34.0 & 241.0 &  62.1\\
  M3-GPU & GPU parallel & MP & mem &   12.4 &   5.1  & mem &   2.4  &  1.9  &  0.0  &  0.0\\
  \rowcolor{green!20}
  X86-arch & arch-aware & SB & arch-aware & 48.3 & 2.9 &  arch-aware & 1.1 & 2.5 & N/A & N/A & \\
  \hline
\end{tabularx}
\end{table*}


\papernameshort can be used to implement a covert channel in the following way: a process receives communication by running a litmus test and recording the number of observed re-orderings. 
A time-series analysis is run to decode the signals into \textit{high} ($\uparrow$), \textit{low} ($\downarrow$), and \textit{space} ($\emptyset$) signals. To send data, the process executes different stress patterns which encode the $\uparrow$ and $\downarrow$ signals, while a pause in stress encodes $\emptyset$.


%

To ease implementation, we focus on a single direction channel with one sender and receiver process. The receiver utilizes a \textit{test configuration} (litmus test and testing framework), while the sender utilizes two types of stress configurations to encode the two signals, which are chosen based on the results from the fuzzing campaign. 
We summarize the configurations in \mytab\ref{tab:covert-configs}. We see that the MP and R litmus tests are the most sensitive; each of the different testing frameworks and stress techniques are used, except the perpetual testing framework. We note that the X86-arch row in the table corresponds to early experiments on exploiting low-level system details in \papernameshort and is described more thoroughly in \mysec\ref{sec:arch-aware}.



To decode signals, the receiver process must both \textit{classify} signals it receives and \textit{transition} between signal states.

\paragraph{Signal Classification}

We found that the number of re-orderings with a reliable stress configuration is approximately normally distributed, allowing us to calculate the mean and standard deviation over a set of samples.
The receiver maintains a window of test samples to account for the natural variability of weak behaviors and lack of synchronization between the sender and receiver.

The sender transmits signals by running the $\uparrow$, $\downarrow$, and $\emptyset$ signals. 
The receiver calculates the likelihood of each sample in the window coming from a given signal distribution using a t-test parameterized by the signal's mean and standard deviation. 
Each sample in the window is then classified as the signal with the distribution that most closely matches the sample by ranking the p-values of the t-test results in descending order. 
The receiver classifies a window as the signal matching the classification of the majority of the samples in the window.
While this approach works well for CPUs, we found that on the M3-GPU the number of memory re-orderings during a $\emptyset$ signal is usually 0 and not normally distributed. Therefore, we use a cutoff heuristic on this GPU and classify a sample as a $\emptyset$ signal if the number of re-orderings is less than the cutoff. 


\paragraph{State Transition}

The receiver implements a state machine to transition between signals. It starts in a \textit{standby} state, signifying that it is waiting to see a $\uparrow/\downarrow$ signal. Once it has enough samples to fill a window, the receiver classifies the window and transitions to either the $\uparrow$ or $\downarrow$ state. The receiver stays in the state until it classifies a window as the $\emptyset$ signal, at which point it records a bit and returns to the \textit{standby} state. On the M3-GPU, we observed that a $\uparrow$ signal is sometimes misclassified, so we include a \textit{$\downarrow'$} state which the receiver transitions to before transitioning to either the $\uparrow$ or $\downarrow$ state.



\begin{table}
\centering
\caption{\papernameshort Covert channel metrics. The X86-arch system is highlighted because it is designed differently to show-case the potential of \papernameshort if low-level system details can be exploited (see \mysec\ref{sec:arch-aware})} 
\label{tab:covert-results}

\vspace{.1cm}

\begin{tabular}{l r r r}
\toprule
\textbf{Device} & \textbf{Window Size} & \textbf{Accuracy} & \textbf{Bits/second} \\
\midrule
 X86    & 5 & 98\% & 0.32\\ 
 Arm    & 5 & 94\% & 0.36 \\
 M1     & 5 & 95\% & 0.66 \\
 M3-GPU & 3 & 95\% & 16.05 \\
\rowcolor{green!20}
 X86-arch & 5 & 94\% & 29448.90 \\
 \hline
\end{tabular}
\end{table}

\paragraph{Experiments and Results}

We test the accuracy and speed of the covert channel by sending random bit strings from the sender to the receiver. The accuracy of the channel can be computed using the Levenshtein distance, i.e., the number of single-bit edits (insertions, deletions, or swaps) between the reference string and the received string. We attain a percentage by dividing the distance by the length of the reference string. 

\mytab\ref{tab:covert-results} summarizes our results, showing the highest average bits per second (bps) we were able to achieve on the covert channel on different devices while maintaining >90\% accuracy. The results were obtained by sending random 13-character ASCII strings (104 bits) 10 times and calculating the average speed and accuracy across the trials.

On the M3-GPU, we can achieve an average accuracy of 95\% with a speed of 16 bps.
Conversely, we see a much lower bps on CPUs due to system noise from other processes on the OS, requiring more test iterations and a larger window size for increased reliability.
These two factors mean that the transmission rate is 2 orders of magnitude slower than the GPU channel in order to reach a similar accuracy. We observe that the M1-CPU covert channel is roughly $2\times$ as fast as the Arm and X86; this is because the M1-CPU had access to the thread launch stress, which, recall from \mysec\ref{sec:fuzzing-results}, 
yields reliable high signals. Because Arm and X86 operate across KVM boundaries, thread launch stress is less reliable and thus, they use memory stress. Lastly, X86-arch (described in \mysec\ref{sec:arch-aware}) uses specialized stress that exploits low-level details and achieves several orders of magnitude higher bps.

In relation to prior work, \cite{sophia-out-of-order} implements a covert channel for X86 based on re-orderings achieving roughly 1 bps while utilizing a timer. Other timerless approaches achieve bps rates similar to ours~\cite{wang2022hertzbleed,Lipp2022AMDPA,9519416}. Finally, highly specialized timerless approaches can achieve kilobits per second~\cite{yu2023synchronization}; we show that \papernameshort has the potential for this speed in X86-arch, described more in~\mysec\ref{sec:arch-aware}.

\subsection{\papernameshort Fingerprinting \label{sec:fingerprinting}}
For application fingerprinting, we cannot rely on fine-tuned stress signals from our fuzzing results.
However, as illustrated in \mysec\ref{sec:fingerprinting-intro}, we show that certain application classes have distinguishable signals that can be detected by \papernameshort. 

\paragraph{Deep Neural Networks Architectures}
Different DNNs can vary widely in their \textit{architectures}, i.e., they can differ by layer operation, depth, memory usage, and computational intensity.
These architectural differences manifest in their underlying implementation on a given platform or accelerator, resulting in distinct memory access patterns.
These differences are akin to the different stress signals we search for in our fuzzing campaign, especially related to memory stress.
Thus, a \papernameshort fingerprinting attack may be able to distinguish different DNN architectures running on different processes. 

\begin{figure}
    \centering
    \includegraphics[width=\linewidth]{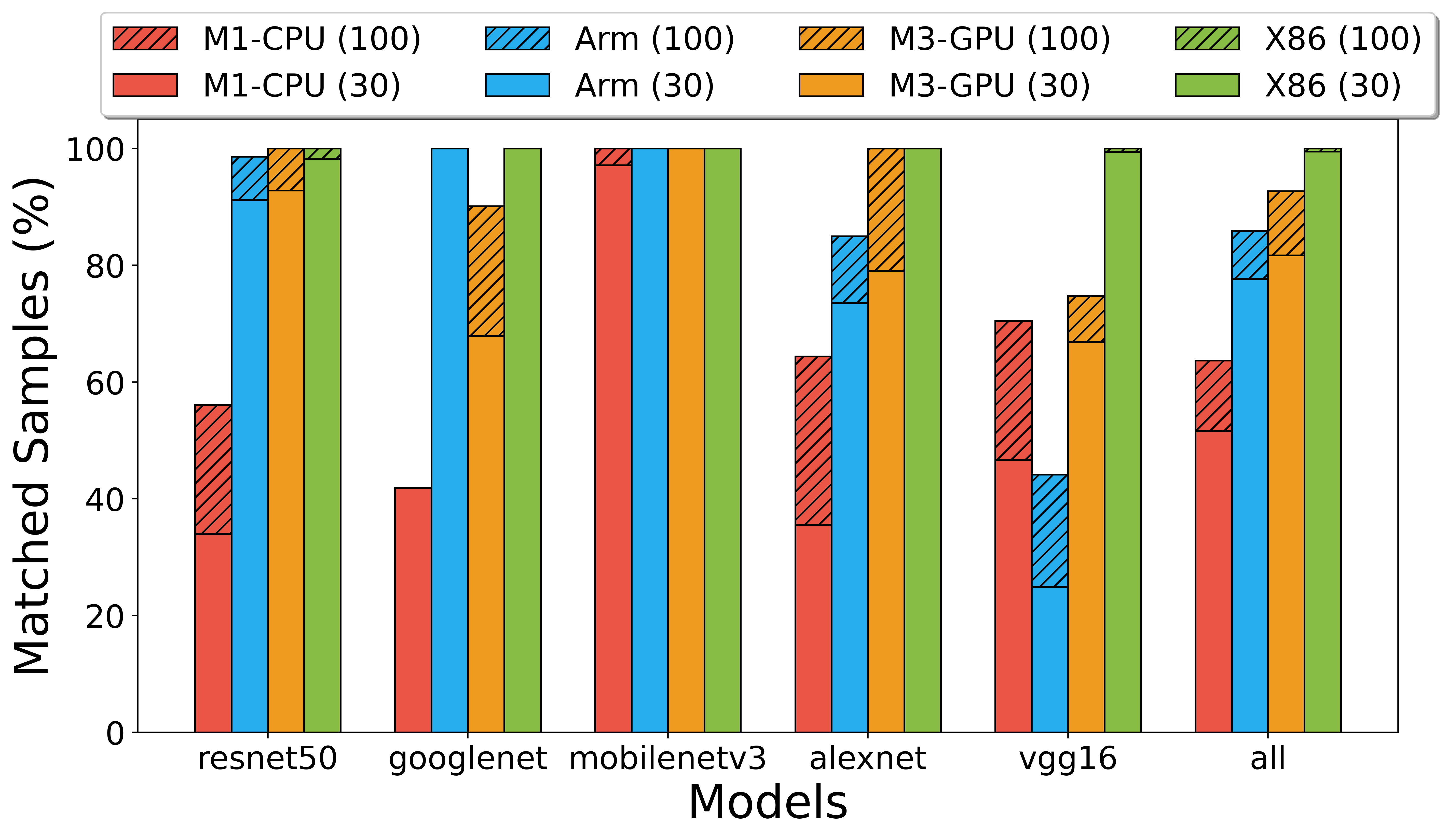}
    \caption{Results of classifying DNN architectures. This graph shows the percentage of matched samples using an independent sample t-test against a training set of memory re-ordering observations with sample sizes of 30 and 100.}
    \label{fig:dnn-fingerprinting}
\end{figure}

In these experiments, we select 5 common DNN architectures, shown in \myfig\ref{fig:dnn-fingerprinting}. 
To select a test for the attack process, we again consult our fuzzing data, looking for a configuration that is reliably sensitive to memory stress.
The attack process then collects test samples while the victim simultaneously runs DNN inference.


The details of our experiment are the following: let $A$ be the set of candidate models we wish to classify, let $s$ be the system we are testing (e.g., Arm, X86, M1-CPU, or M3-GPU), and let $t$ be the litmus test determined to be sensitive to memory stress on $s$.
For each DNN architecture $d \in A$, we collect 4K observations of executing $t$ on $s$ while running $d$ on a different process. We designate the first 2K observations to be the training set $S_d$, and the second 2K to be the test set $T_d$. 
For a DNN architecture $d$, we then randomly select $z$ observations from $T_d$ into sample $t$. 
We then compare $t$ against each training set $S_i$ for $i \in A$ using an independent samples t-test, classifying $t$ to whichever DNN provided the best fit.


We run these experiments across two different sample sizes: small (30) and large (100). 
\myfiglong\ref{fig:dnn-fingerprinting} shows the results of running 1000 trials per sample size, with each trial sampling from a randomly chosen test set $T_d$. 
As sample size increases, classification accuracy improves, reaching over 80\% at a size of 100 on most devices.
We tune the litmus test iterations such that 30 samples takes less than 5 seconds on each device.

The mobilenetv3\_small and alexnet architectures were easily classified across all devices, while vgg16 had comparably lower accuracy on every device except the X86 CPU.
Indeed, the distributions of vgg16 and resnet50 shown in \myfig\ref{fig:intro-histogram} are relatively close on the M3-GPU. We find that the M1-CPU has the lowest accuracy, however, pilot experiments showed that increasing the litmus test iterations (as opposed to the number of samples) appeared to increase its accuracy substantially. Thus, this attack requires tuning across multiple dimensions, but high accuracy can be obtained across these devices.
Furthermore, utilizing more complex classification techniques such as those in \cite{gao2024deep} could help to further refine the accuracy of DNN classification using \papernameshort.


\paragraph{Launching Applications}
Given that our fuzzing results show a high impact for thread launch stress, we now explore if \papernameshort can be used to fingerprint other types of system behavior. 
We select the M1-CPU device since it is especially sensitive to this stress type.

We select an application that might be conceptually similar to our thread launch stress, where threads are repeatedly launched and joined. For this, we opted to use Google Chrome, as it is heavily multithreaded. We explore if \papernameshort can be used to identify when an application like Google Chrome is opened or closed. To do this, we execute a script that repeatedly opens and closes Chrome, with 3 seconds between launching and closing and 2 seconds between re-launching.

\begin{figure}
    \centering
    \includegraphics[width=1\linewidth]{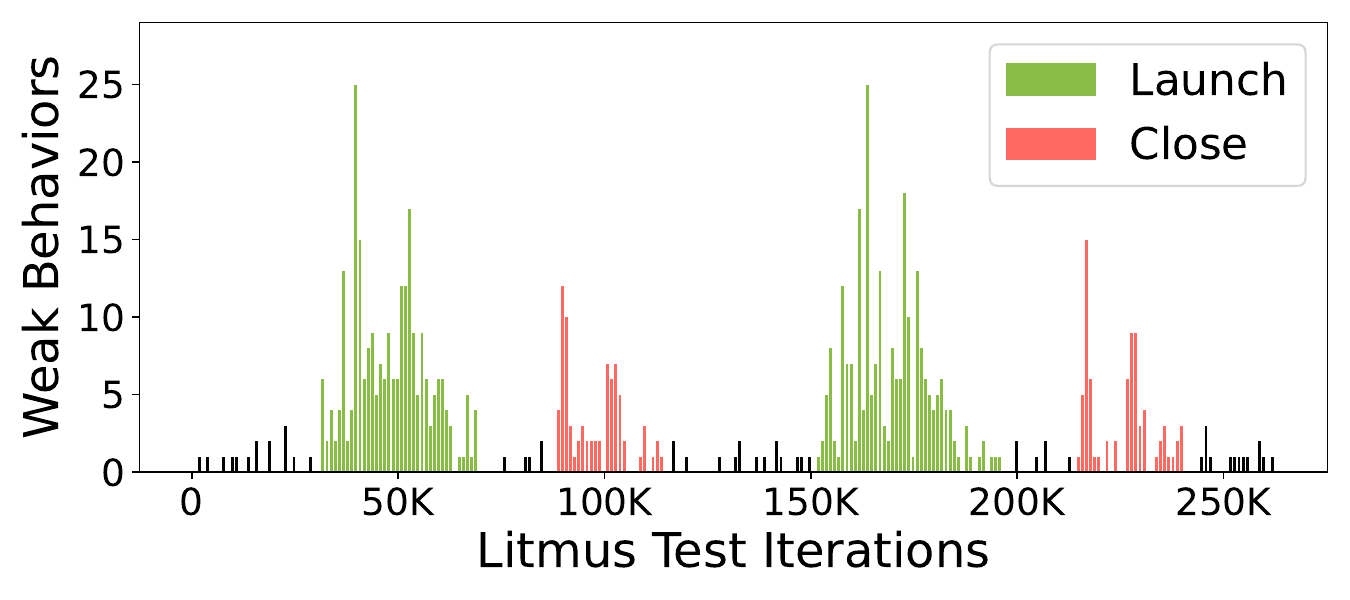}
    \caption{Memory re-orderings time series data across launching and closing Google Chrome on the M1-CPU.}
    \label{fig:google-chrome}
\end{figure}

The attacker uses the Read litmus test in the basic testing framework, as that happens to be the most sensitive, and constantly samples the number of re-orderings observed at intervals of 1k iterations. \myfiglong\ref{fig:google-chrome} shows the result of this experiment, with the x-axis representing timestamps and the y-axis showing the number of observed re-orderings. We see very clear patterns when Google Chrome is launched (the longer and taller region of re-orderings, colored in green), when it is closed (shorter regions of re-orderings, colored in red), and when the system is idle (lower re-orderings, colored in black). This shows the potential for \papernameshort to be used to fingerprint other application behavior, especially related to interactive applications on consumer devices.

Our pilot experiments show that many types of system behavior increase the number of re-orderings observed on M series Apple devices. For example, even swapping tabs on a terminal or launching an application dramatically increased the number of observed re-orderings. Because the low-level cause of these re-orderings has not yet been determined, it is difficult to appreciate their full potential or offer rigorous mitigation approaches. Given this, we believe future works will be able to increase the fingerprinting precision and impact of \papernameshort .

\subsection{Targeted Architectural Attacks}
\label{sec:arch-aware}

Up to this point, our attacks require no low-level architectural knowledge, instead using our empirical fuzzing results. 
Now, we show early results on the potential of \papernameshort attacks which exploit low-level hardware details. We show this on our X86 processor, as it was the most straightforward platform to run experiments on and X86 is typically well-documented. Our experiments are in the context of a covert channel, where we designed a custom sender and receiver framework called \textit{arch-aware}. We target the performance cores of this device, where each core can execute two hardware threads (via hyperthreading) and has its own 12-way set associative L1 cache of size 48KB with a cache line size of 64 bytes.

Our hypothesis is that the X86 store buffer (responsible for re-orderings) would be sensitive to whether the L1 cache contains the \texttt{x} or \texttt{y} location in the litmus test. The receiver (which implements the litmus test) is mapped to two distinct cores. We found that by evicting one of the locations from the L1 (say \texttt{x}), while ensuring the other location (say \texttt{y}) remained in the L1 provided extremely high and reliable signals (i.e., re-orderings were observed in over 90\% of iterations). 

To implement a covert channel in this framework, we mimic a sender in another thread which is mapped to the same core as thread 0 in the receiver. To provide a signal, the sender simply writes to 12 locations (the size of a cache set) that map to the same cache set as memory location \texttt{x}, thus evicting \texttt{x} from thread 0's L1 cache. \mytablong\ref{tab:covert-configs} shows that these signals are very high and very reliable. 

Extending this method, we are able to monitor multiple cache sets: our X86 has 64 L1 cache sets, which allows us to instantiate 63 litmus tests. The \texttt{x} locations map to different cache sets, while the \texttt{y} locations all map the 64th cache set, as it should not be evicted. These signals are so clear that the tests only need to execute for 15 iterations. Furthermore, given that multiple bits are sent at a time, there is no need for the $\emptyset$ signal as the first bit can act as a virtual clock, modulating high and low between signals. 
This highly tuned configuration is able to achieve nearly 30k bps, several orders of magnitude higher than other approaches (see X86-arch in \mytab\ref{tab:covert-configs} and \ref{tab:covert-results}).

\paragraph{Towards a Full Attack} Our results from X86-arch are preliminary and are meant to show the potential of \papernameshort when exploiting low-level system details. We were unable to fully implement arch-aware as a full attack, as the re-orderings seemed sensitive to more than simply cache sets. We found that signals degraded in unpredictable ways as we tried to partition the memory more fully across processes; and conversely, strong reliable signals showed up in other configurations that did not seem to correspond to any system documentation we were aware of. Thus, we believe more detailed investigations are necessary, and once the attack is more understood, then it would likely be possible to fully implement high bandwidth covert channels, as well as more fine-grained data extraction attacks, e.g., cryptographic key extraction~\cite{yu2023synchronization}.




\section{ \papernameshort Mitigations and Remediations \label{sec:mitigations}}

Similar to other side-channel attacks, e.g., prime-and-probe, mitigations to \papernameshort are difficult and invasive. Furthermore, unless vendors release precise explanations for memory re-orderings (unlikely), then mitigations will be speculative or empirical. For current systems, we see two mitigation paths:

\paragraph{Disallowing Memory Re-orderings} 
Program analysis techniques have been used to disallow memory re-orderings by automatically inserting \textit{fence} instructions. To mitigate \papernameshort using this approach, a system would have to force all untrusted programs to be compiled in a way that removes re-orderings, and disallow programming features that bypass compiler analysis, e.g., inline assembly.



A naive approach can remove all re-orderings by placing a fence after every memory instruction; however, this disallows many hardware optimizations, and as such, has a high performance overhead (as reported in~\cite{10.1145/2994593, 10.1145/2908080.2908114}). Other approaches perform more complex analysis to identify memory accesses that are potentially shared across threads to prune the number of fences. The overhead of these approaches is much less, reportedly around $1.5\times$ on average~\cite{10.1145/3314221.3314611}. Some languages follow a paradigm called \textit{sequential consistency for data-race free programs} (abbreviated as SC-DRF)~\cite{10.1145/325164.325100}. Languages that are SC-DRF guarantee that if programs follow certain rules (i.e., they do not have data races) then no memory re-orderings will be observable. If a system strictly enforced SC-DRF programs (such as in safe Rust), then memory re-orderings (and, thus, \papernameshort) would not be possible to observe.


\paragraph{Signal Obfuscation}

A common side-channel mitigation is to 
obfuscate compromised execution characteristics. For example, a memory-based attacks can be mitigated if the victim ensures input-oblivious memory accesses~\cite{10.1007/11605805_1}.  Other approaches might insert random memory accesses in such a way that the information leaked is no longer useful. Prior work has proposed a variety of automated approaches that accomplish this, from compiler techniques~\cite{7987067}, to transparent memory management techniques~\cite{zhou2016software}.

We show two characteristics that leak information through \papernameshort. The first is memory access patterns, which can likely be mitigated similar to other memory side-channels. The second is thread launching behaviors, which are novel for \papernameshort. Applications (e.g., Google Chrome) would need to modify their thread management logic, e.g., to be uniform with other applications. Given the complexity of large software, combined with the complexity of parallel programming, such mitigations will be difficult and invasive. 
Our pilot experiments also found that \papernameshort seems to be sensitive to a variety of other system behavior (e.g., rapidly switching terminal tabs). To implement a robust mitigation, the cause of these re-orderings would need to be more fully understood.

\paragraph{Online Detection} 
A fingerprinting attack requires only two cores to execute the litmus test, and <2MB of memory.
The overhead of running a \papernameshort in the background varies across systems and applications, but we found that for DNN inference, \papernameshort listeners caused only a 25\% slowdown on Apple CPUs. This overhead reduces for systems with more parallelism, e.g., for the M3-GPU, the overhead of \papernameshort listeners was negligible.  Thus, it is unlikely a resource watchdog would flag a \papernameshort attack as being an outlier. 

\section{Related Work}

\paragraph{Memory-based Side-channels}
Many recent works have exploited a variety of different memory mechanisms as side-channels; 
For example, 
\cite{jiang2024sync} implements a side-channel by measuring contention on storage devices on Linux systems. They show how this can be used to implement a covert channel and do coarse-grained application fingerprinting.
Other works analyze implicit (and undocumented) memory compression techniques of different GPUs~\cite{gpu-zip}; the memory traffic produced by the compressed data can be used to reconstruct images. Lastly, X86 prefetch instructions were shown to leak cache coherence states~\cite{advasarial-prefetch}, which can be used to implement transient execution attacks more efficiently than prior works. 

\papernameshort distinguishes itself in two significant ways: (1)~\papernameshort is lower capability, requiring no timers, no hardware monitors, and no detailed reverse engineering for low-precision configurations; (2)~\papernameshort exists across many devices, whereas prior works are usually specialized.


\paragraph{Memory Model Testing and Modeling} Some prior works have explored causes and characteristics of re-orderings.
For example,~\cite{tso-bounding} uses empirical testing to determine the size of the X86 store buffer; other works showed how memory bank conflicts cause re-orderings on GPUs~\cite{asplos15}. These low-level details could greatly increase the precision of \papernameshort attacks (as shown in \mysec\ref{sec:arch-aware}). However, they seem to exist only for certain behaviors on (older) architectures.

Other work has modeled microarchitectural features that cause re-orderings (such as cache protocols)~\cite{7856585}; utilizing these models, they are able to validate conformance to higher-level specifications. Similarly, another work used happens-before relational models to capture many types of microarchitectural side-channels~\cite{8574598}. While this level of modeling could yield insights into the cause of memory re-orderings, it requires significant low-level knowledge, much of which is proprietary, and as such, mainstream commercial systems have not been modeled in this way. 

\paragraph{DNN Architecture Fingerprinting} Other works have used side-channels to extract/predict DNN architectures, similar to how \papernameshort is used in \mysec\ref{sec:fingerprinting}. For example, \cite{yan2020cache} use Prime+Probe and Flush+Reload cache attacks to learn sizes and number of matrices computed with high-performance GEMM libraries.
Other works use power consumption as a side-channel in order to carry out a model extraction attack~\cite{xiang2020open}. 
More sophisticated model extraction attacks based on power consumption were shown in~\cite{gao2024deep}, which recovered network structure by analysing the time-dependent energy trace. These works are highly specialized to certain devices and require detailed timers or low-level power monitoring APIs, whereas \papernameshort is low capability and shown to be widely applicable across many different devices.

\section{Conclusion}

We present a novel, timerless side-channel attack utilizing memory re-orderings called \papername. 
We show that this attack impacts most mainstream processors, including Arm, X86, and Apple CPUs, as well as NVIDIA, AMD, and Apple GPUs. 
We show the potential for \papernameshort to be used in classic attacks such as covert channels and application fingerprinting. If future work is able to more precisely identify the cause of memory re-orderings (possibly focusing on a single processor), it will likely enable more targeted attacks, as well as inform robust mitigation techniques.


\section{Responsible Disclosure \label{sec:ethics}}



We provided a pre-print and summary of this work to the security teams at the following companies: Apple, Amazon, AMD, ARM, Google, Intel, Microsoft, NVIDIA, Qualcomm, and Samsung. All acknowledged the finding and approved of the publishing timeline.

\bibliographystyle{plain}
\bibliography{ref}

\end{document}